\documentclass[epj]{svjour}

\usepackage{graphicx}
\usepackage{fancyhdr}
\def\makeheadbox{{
 }}
\def\mytitle{Complementarity of LHC and ILC}
\def\myauthors{S.Y. Choi}
\def\mytype{Plenary}
\def\mysession{S.Y. Choi}

\begin{document}
\title{Complementarity of LHC and ILC}
\author{S.Y. Choi
\thanks{\emph{Email: sychoi@chonbuk.ac.kr}}
}                     
%
%
\institute{Department of Physics and RIPC, Chonbuk National University,
           Jeonju 561-756, Korea}
%
\date{}
\abstract{Two next-generation high-energy experiments, the Large Hadron Collider
          (LHC) and the $e^+e^-$ International Linear Collider (ILC),
          are highly expected to unravel the new structure of matter and forces
          from the electroweak scale to the TeV scale. In this talk we give a
          compelling but rather descriptive review of the complementary role of
          LHC and ILC in drawing a comprehensive and high precision picture of
          the mechanism breaking the electroweak symmetries and generating mass,
          the unification of forces and the structure of spacetime. Supersymmetry
          is exploited in this description as a prototype scenario of
          the physics beyond the Standard Model.
\PACS{{12.60.-i} {Models beyond the standard model} --
      {12.60.Jv} {Supersymmetric models}}
} 
\maketitle
\section{Introduction}
\label{intro}

Particle physics has been very much successful in unraveling the basic laws of
nature at the smallest accessible length scale and it has revealed
a consistent picture, the Standard Model (SM), adequately describing the structure
of matter and forces. However, many theoretical
arguments and experimental observations strongly indicate that the model is
incomplete and it should be embedded in a more fundamental theory, addressing
a set of crucial questions to be approached experimentally
at the TeV scale (Terascale): the mechanism of electroweak
symmetry breaking (EWSB) and mass generation; the unification of forces, including
gravity finally; and the structure of spacetime. This set of particle physics
questions is intriguingly connected to the cosmology questions such as the
nature of particles comprising cold dark matter (CDM) and the origin of the
baryon asymmetry in the universe.

The next generation of high-energy accelerators will get access to the
Terascale with a high expectation of providing decisive answers to these
questions \cite{R0a,R0b}. LHC with a c.m. energy of 14 TeV \cite{R1a,R1b,R1c}
will put the first springboard in 2008 for breakthrough discoveries in the
EWSB sector and in the physics beyond the SM (BSM). However, the processes and
the detections of new physics at LHC are extremely complicated. Therefore, a lepton
facility with clean environments (and, if possible, with various
facets) is required to complement this hadron machine
in drawing a comprehensive and high-resolution picture of EWSB and of the BSM.
ILC \cite{R2a,R2b,R2c,R2d,R2e,R2f}, which is now in the design phase, can be
an excellent counterpart to LHC. The ILC energy of 500 GeV in the first phase
and 1 TeV in the upgraded phase in the lepton sector is equivalent in many
aspects to the higher LHC energy of about five TeV in the quark sector.
Moreover, ILC covers one of the most crucial energy ranges including the
characteristic EWSB scale ($v=246$ GeV). [If the BSM scale revealed
at LHC might be beyond the reach of ILC, it could be accessed later by a
multi-TeV $e^+e^-$ collider such as the Compact Linear
Collider (CLIC) \cite{R3}.]

There exist several world-wide studies of the LHC and ILC
interplay \cite{R4a,R4b}. In particular, the LHC/ILC Study Group, formed as
a collaborative effort of the hadron and lepton collider experimental
communities and theorists, has completed a very comprehensive working group
report with many detailed studies of various conceivable BSM scenarios \cite{R4a}.
This talk will not cover all the topics unlike the report but it will give
a compelling but descriptive review of the complementary role of LHC and ILC in
drawing a model-independent and high-resolution picture of the new Terascale
physics and revealing the fundamental theory at the scale close to the grand
unification (GUT) or Planck scale. Supersymmetry (SUSY) will exclusively be
considered as a BSM prototype concept in this description. For an excellent
recent review for alternative scenarios, see, for example, Ref.\,\cite{R5}.

\section{Supersymmetric path}
\label{sec:2}

In supersymmetric theories a light Higgs boson is generated and the electroweak
(EW) scale is stabilized naturally against the GUT/Planck-scale background.
The presence of the supersymmetric particle spectrum is essential
for a high-quality unification of the three SM gauge couplings at a high energy
scale \cite{R6a,R6b,R6c,R6d}. It offers a natural CDM candidate. Moreover, local
SUSY provides a rationale for gravity by demanding the existence of massless
spin-2 gravitons. In short, if realized in nature, SUSY will have an impact
across all microscopic and cosmological scales.

\begin{figure}
\begin{center}
\includegraphics[width=0.28\textwidth,height=0.28\textwidth,angle=0]{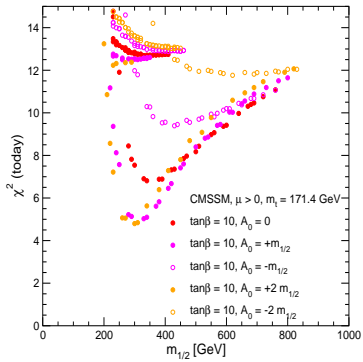}
\end{center}
\caption{The combined $\chi^2$ function for electroweak precision observables
         and $B$-physics observables; Ref.\,\cite{R9}.
         }
\label{fig1}       
\end{figure}

\begin{figure}
\begin{center}
\includegraphics[width=0.28\textwidth,height=0.26\textwidth,angle=0]{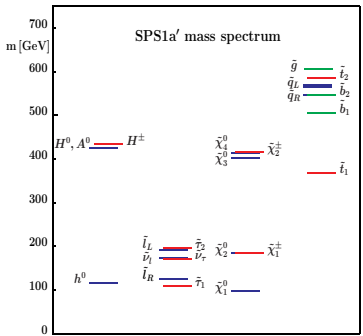}
\end{center}
\caption{Mass spectrum of supersymmetric particles and Higgs bosons in the
         reference point SPS1a$^\prime$; Ref.\,\cite{R10}.}
\label{fig2}       
\end{figure}

There is no firm prediction yet for the SUSY mass scale. However,
there are important direct constraints on the mass scale due to the absence of
sparticles at LEP and the Tevatron, and also indirect constraints
from the LEP lower limit of 114 GeV on the Higgs mass \cite{Higgs_bound},
the reasonable agreement of SM calculations of $b\to s \gamma$ \cite{bsr1,bsr2},
the BNL measurement of the anomalous magnetic moment of the muon
$a_\mu$ \cite{R7a,R7b}, and also from the measurement of the CDM density
at WMAP \cite{R8}. As shown in Fig.\,\ref{fig1}, a global fit to precision EW and $B$-decay
observables indicates a fairly low-mass spectrum for moderate values of the
Higgs mixing parameter $\tan\beta$ in the constrained minimal supersymmetric
SM (CMSSM) \cite{R9}. In the favorable case several non-colored supersymmetric
particles such as lighter neutralinos and sleptons should be observed at ILC
in the first phase with its c.m. energy of 500 GeV and even the heavier
non-colored particles and the lighter top squark in the upgraded
phase with its c.m. energy of 1 TeV. The spectrum corresponding to a parameter
set with close to maximal possibility is depicted in Fig.\,\ref{fig2}.
This spectrum had been chosen as a benchmark set for a minimal
supergravity scenario in the SPS1a$^\prime$ project \cite{R10}.

LHC and ILC can provide us with a perfectly combined tool for exploring
SUSY \cite{R4a}. The heavy colored supersymmetric particles, squarks and
gluinos, can be discovered for masses up to 3 TeV with large rates at LHC.
The properties of the potentially lighter non-colored particles, charginos,
neutralinos, sleptons and Higgs bosons, can be studied very precisely at
ILC by exploiting, in particular, beam polarizations. Once the properties
of the light particles are determined precisely at ILC, the heavier particles
produced at LHC can subsequently be studied in the cascade decays with
much greater precision. Based on the coherent LHC and ILC analyses we can then
take the following supersymmetric path by
\begin{itemize}
\item[{$\bullet$}] measuring the masses and mixings of the newly produced
      particles, their decay widths and branching ratios, their production
      cross sections, etc;
\item[{$\bullet$}] verifying that there are indeed the superpartners of the SM
      particles by determining their spin and parity, gauge quantum numbers
      and their couplings;
\item[{$\bullet$}] reconstructing the low  energy Lagrangian with the smallest
      number of assumptions, i.e. as model independently as possible;
\item[{$\bullet$}] and unraveling the fundamental SUSY breaking mechanism and
      shedding light on the physics at the very high energy (GUT or Planck) scale,
\end{itemize}
from the EW scale to GUT/Planck scale, on one side, for the reconstruction
of the fundamental SUSY theory near the Planck scale and, on the other side, for
the connection of particle physics with cosmology.

\section{Higgs bosons}
\label{sec:3}

In SUSY theories the Higgs sector includes at least two iso-doublet
scalar fields so that at least five more physical particles are
predicted \cite{Higgs_LHC}. In the minimal supersymmetric SM (MSSM) the mass of
the lightest neutral scalar Higgs particle $h$ is bounded above to about 140 GeV,
while the masses of the heavy neutral scalar and pseudoscalar Higgs bosons,
$H$ and $A$, and the charged Higgs bosons, $H^\pm$, may range from the EW
scale to a multi-TeV scale. The upper bound on the lightest Higgs mass is relaxed
to about 200 GeV in more general scenarios if the fields remain weakly interacting
up to the Planck scale.

\subsection{MSSM Higgs Bosons}
\label{sec:3-1}

While the light Higgs boson $h$ can be detected at LHC in the full range of the
$M_A$ and $\tan\beta$ parameter space, the heavy Higgs bosons cannot be detected
in a wedge centered around the medium mixing angle $\tan\beta\sim 7$ and opening
from the masses of about 200 GeV up to higher values \cite{R1a,Heavy_Higgs}.
This region can however be covered considerably by ILC.

\begin{figure}
\begin{center}
\includegraphics[width=0.3\textwidth,height=0.3\textwidth,angle=0]{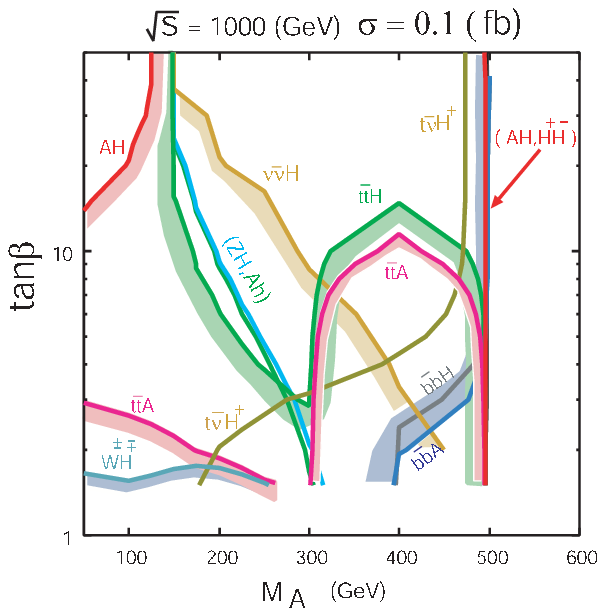}
\end{center}
\caption{Cross section contours of various heavy MSSM Higgs production
         processes in the $M_A/\tan\beta$ plane for $\sqrt{s}=1$ TeV;
         Ref.\,\cite{R11}.}
\label{fig3}       
\end{figure}

At ILC the search and study of the light Higgs boson $h$ follows the pattern
very similar to the SM Higgs boson in most of the parameter space and the
heavy Higgs bosons are produced in mixed pairs at ILC:
$ e^+e^-\to HA\ \ \mbox{and}\ \ H^+H^-$.
Therefore, the wedge can be covered by pair production in $e^+e^-$ collisions
for masses $M_{H,A}\leq \sqrt{s}/2$, i.e., up to 500 GeV in the TeV phase of
the ILC machine, cf. Fig.\,\ref{fig3} \cite{R11}. Moreover, single production in
photon-photon collisions, $\gamma\gamma \to H$ and $A$,
can cover the wedge up to Higgs masses of 800 GeV if a fraction of 80\% of the
total $e^+e^-$ energy is transferred to the $\gamma\gamma$ system by
Compton back-scattering of laser light \cite{R12}.

After the Higgs particles are discovered, it must experimentally be established
that the Higgs mechanism is responsible indeed for breaking the electroweak
symmetries and for generating the masses of the fundamental SM particles. This
requires profiling the Higgs bosons precisely. First model-independent analyses
of the properties can be performed at LHC by measuring the Higgs masses,
the ratios of some Higgs couplings and the bounds on couplings \cite{R13}.

However, the truly model-independent and  high-resolution determination of the
profile of the light Higgs boson $h$ -- the mass, the spin of the particle, its
couplings and the trilinear self coupling --  can be made at ILC with the clear
signals of Higgs events above small backgrounds in the processes such as
Higgs-strahlung, $e^+e^-\to Zh$, and $WW$ fusion, $e^+e^-\to \bar{\nu}\nu h$,
and in the process of double Higgs production,
$e^+e^-\to Zhh$ and $\bar{\nu}\nu hh$ \cite{Higgs_LHC}.

\begin{figure}[h]
\begin{center}
\includegraphics[width=0.28\textwidth,height=0.26\textwidth,angle=0]{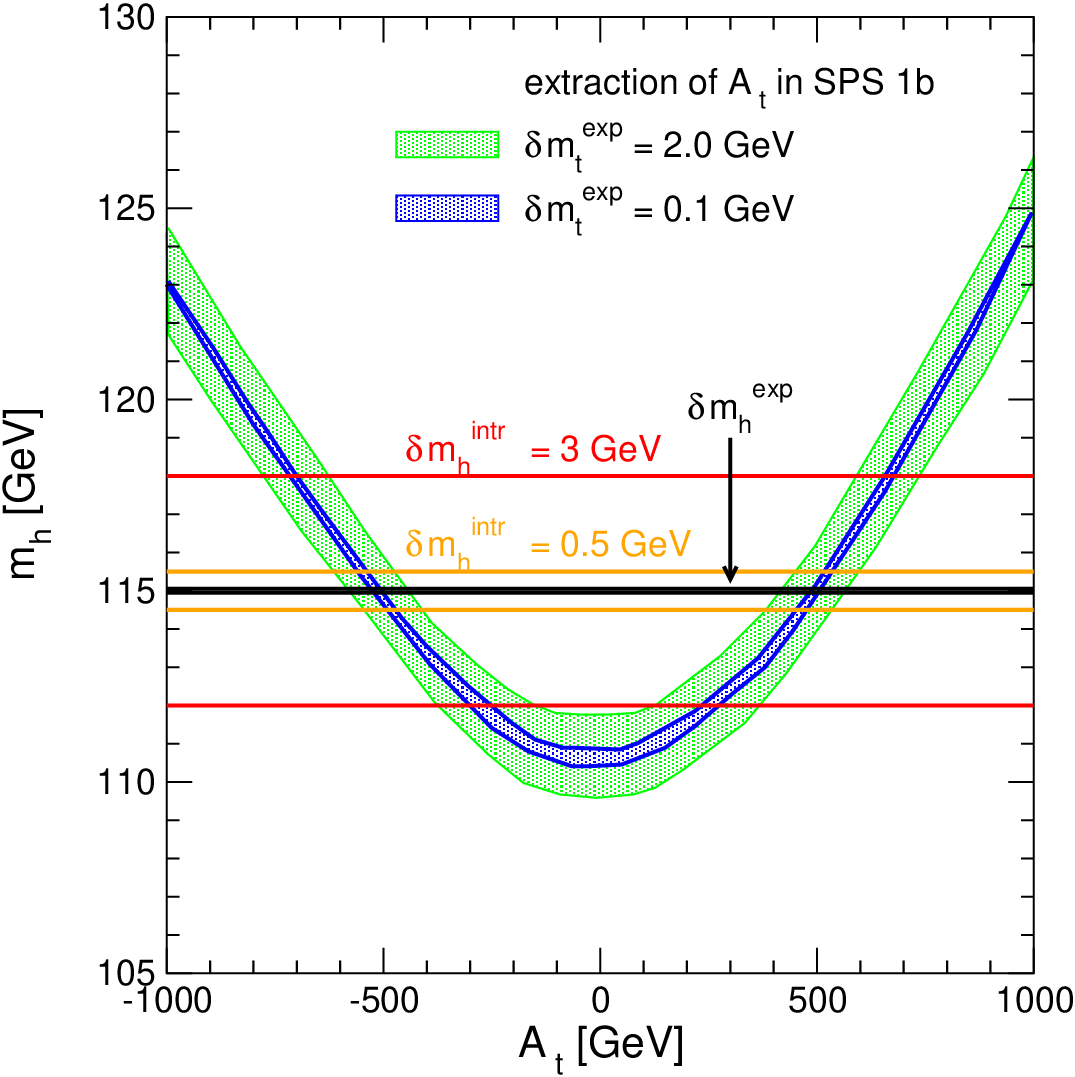}
\end{center}
\caption{Extracting the trilinear coupling $A_t$ from radiative corrections to
         the light MSSM Higgs mass; Ref.\,\cite{R14}.}
\label{fig4}       
\end{figure}

Such high-precision measurements of the light Higgs mass can be exploited to
determine parameters in the SUSY theory which are very difficult to
measure directly. For instance, by evaluating quantum
corrections, the top quark trilinear coupling $A_t$ can be calculated from the
Higgs mass, Fig.\,\ref{fig4}. For an error on the top quark mass of
$\delta m_t=100$ MeV and an error on the Higgs mass of $\delta m_h=50$ MeV,
$A_t$ can be determined at an accuracy of about 10\% \cite{R14}.

\begin{figure}[h]
\vskip 0.0cm
\begin{center}
\includegraphics[width=0.3\textwidth,height=0.3\textwidth,angle=0]{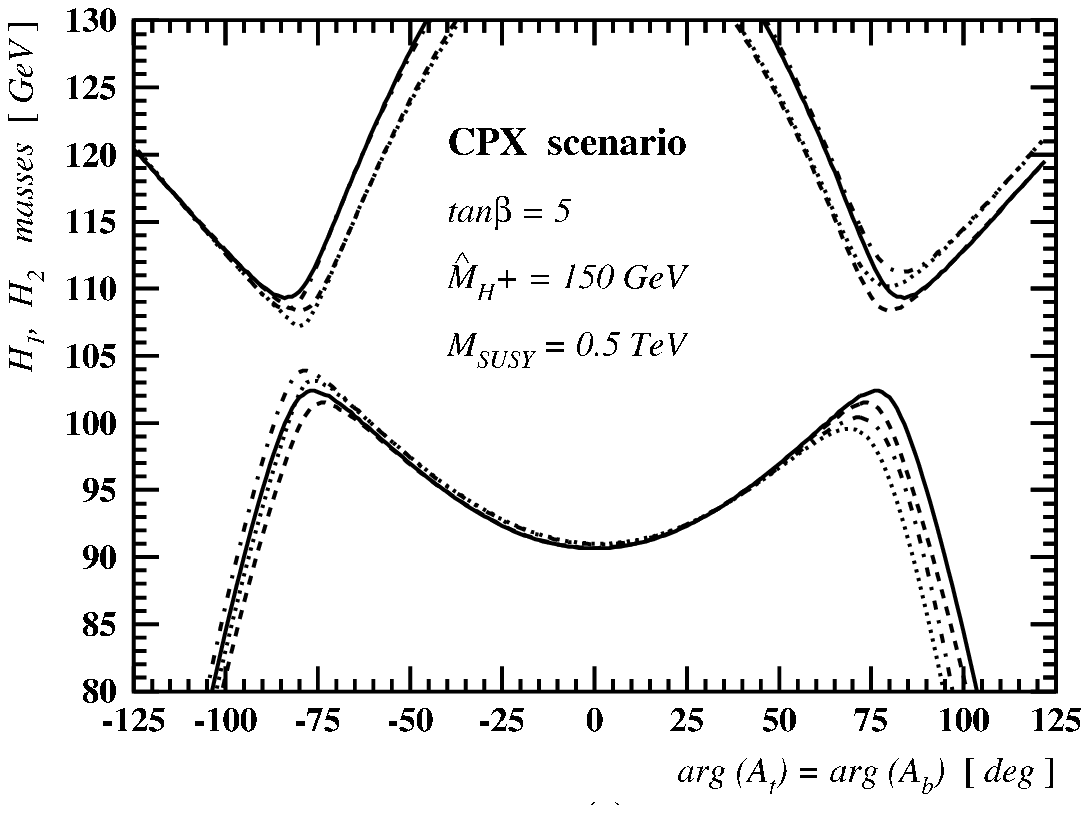}
\end{center}
\caption{Extracting the trilinear coupling $A_t$ from radiative corrections to
         the light MSSM Higgs mass; Ref.\,\cite{R18}.}
\label{fig5}       
\end{figure}

\subsection{CP violation in the MSSM Higgs sector}
\label{sec:3-2}

In the general MSSM \cite{general_MSSM}, the gaugino mass parameters
$M_i$ ($i=1,2,3$), the higgsino mass parameter $\mu$, and the trilinear
couplings $A_f$ can be complex so that they can induce explicit CP violation
in various ways in the model. Their physical rephasing-invariant combinations
affect sparticle
masses and couplings through their mixing, induce CP violating mixing in the
Higgs sector through radiative corrections, influence CP even observables such
as cross sections and also lead to interesting CP odd asymmetries at colliders.
As a result, although stringently constrained by low energy observables like
electric dipole moments (EDMs), the nontrivial CP phases can significantly
influence the collider phenomenology of Higgs and SUSY particles and also the
properties of neutralino CDM \cite{R15a,R15b,R15c}.

\begin{figure}[h]
\begin{center}
\includegraphics[width=0.32\textwidth,height=0.32\textwidth,angle=0]{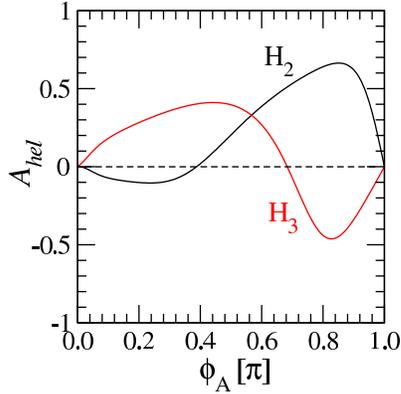}
\end{center}
\caption{The CP-odd asymmetry $A_{hel}$ at the pole of $H_2$ and $H_3$ as a
         function of the rephasing invariant phase $\phi_A$; Ref.\,\cite{R17}.}
\label{fig6}       
\end{figure}

Referring to the CPNSH report \cite{R15a} for an extensive discussion of CP
violation in supersymmetric theories, we simply mention here two examples
of Higgs-sector CP violation. The lightest Higgs boson $H_1$ without definite
CP parity can couple very weakly to the gauge bosons so that the state could have
escaped detection at LEP2 \cite{R16} and the heavy Higgs states $H$ and $A$ can
exhibit CP-violating resonant mixing phenomena when two states are degenerate
in mass in the decoupling regime \cite{R17}. One example of the impact of
the CP-violating Higgs mixing on the Higgs mass spectrum is shown in
Fig.\,\ref{fig5} as a function of the phase of the coupling $A_t$ \cite{R18}.
The other example for studying the CP-violating resonant mixing of
two heavy neutral Higgs bosons is provided by $\gamma\gamma$-Higgs formation
in polarized beams. As shown in Fig.\,\ref{fig6}, the CP violation due to
resonant $H/A$ mixing can directly be probed via the CP-odd asymmetry
$A_{hel}=(\sigma_{++}-\sigma_{--})/(\sigma_{++}+\sigma_{--})$ constructed with
circular photon polarization \cite{R17}.

\subsection{Extended Higgs sector}
\label{sec:3-3}

A large variety of BSM theories such as GUT theories and string theories
suggest extended gauge and Higgs sectors with additional gauge bosons and
Higgs bosons beyond the minimal set of the
MSSM \cite{R19a,R19b,R19c,R19d,R20}.

For example, the next-to-MSSM (NMSSM), the simplest extension of the MSSM,
introduces a complex iso-scalar field, generating a weak scale higgsino mass
parameter $\mu$. The NMSSM Higgs sector is thus extended to include an additional
scalar and a pseudoscalar. In a large area of the parameter space the NMSSM
Higgs sector reduces to the MSSM one but there is a possibility that one of the
neutral Higgs particles, in general the lightest pseudoscalar $A_1$, is very
light enough for the light scalar $H_1$ to decay into pairs of $A_1$ bosons
subsequently decaying to $b$-quarks and $\tau$ leptons, with a large branching
fraction. Nevertheless, a no-lose theorem for discovering at least one Higgs
boson has been established for ILC. The situation is less clear for
LHC \cite{R21a}, although the lightest scalar state $H_1$ may be detected at
LHC via $WH_1$ and $ZH$ production \cite{R21b}.

\section{Supersymmetric particles}
\label{sec:4}

For illustration we adopt the parameters of the minimal supergravity reference
point SPS1a$^\prime$ \cite{R10}. It is characterized by the following values of the soft
parameters at the GUT scale: $M_{1/2}=250$ GeV, $M_0=70$ GeV, $A_0=-300$ GeV,
${\rm sign}(\mu)=+ $ and $\tan\beta=10$ where $M_{1/2}$, $M_0$, $A_0$ and $\mu$
denote the universal gaugino mass, the universal scalar mass, the universal
trilinear coupling and the higgsino mass parameter. The modulus of the higgsino
mass parameter is fixed by requiring radiative electroweak symmetry
breaking \cite{EWSB1,EWSB2,EWSB3,EWSB4,EWSB5} so that $\mu=+396$ GeV. As shown
by the sparticle and Higgs spectrum
in Fig.\,\ref{fig2}, the squarks and gluinos can be studied very well
at LHC and the non-colored charginos and neutralinos, sleptons and
Higgs bosons can be analyzed partly at LHC and precisely at ILC operating
at a c.m. energy up to 1 TeV.

\subsection{Properties of supersymmetric particles}
\label{sec:4-1}

At LHC, the masses can be obtained by analyzing edge effects in the cascade
decay spectra, cf.\,Ref.\,\cite{R22}. An ideal chain is a sequence of two-body decays:
$\tilde{q}_L\to\tilde{\chi}^0_2 q \to \tilde{\ell}_R \ell q \to
\tilde{\chi}^0_1\ell\ell q$. The kinematic edges and thresholds predicted
in the invariant mass distributions of the two leptons and the jet determine
the masses in a model independent way. The determined four particle masses
are used subsequently as input for other decay chains like $\tilde{g}\to
\tilde{b}_1b\to\tilde{\chi}^0_2 bb$ and the shorter chains
$\tilde{q}_R\to q\tilde{\chi}^0_1$ and $\tilde{\chi}^0_4\to\tilde{\ell}\ell$.
However, there are residual ambiguities and the strong correlations between
the heavier masses and the lightest supersymmetric particle (LSP).

\begin{figure}
\includegraphics[width=0.23\textwidth,height=0.26\textwidth,angle=0]{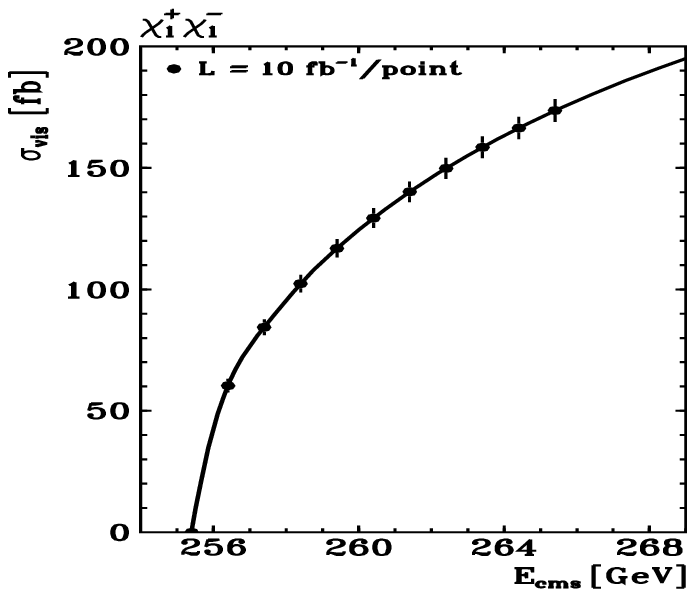}
\hskip 0.5cm
\includegraphics[width=0.21\textwidth,height=0.246\textwidth,angle=0]{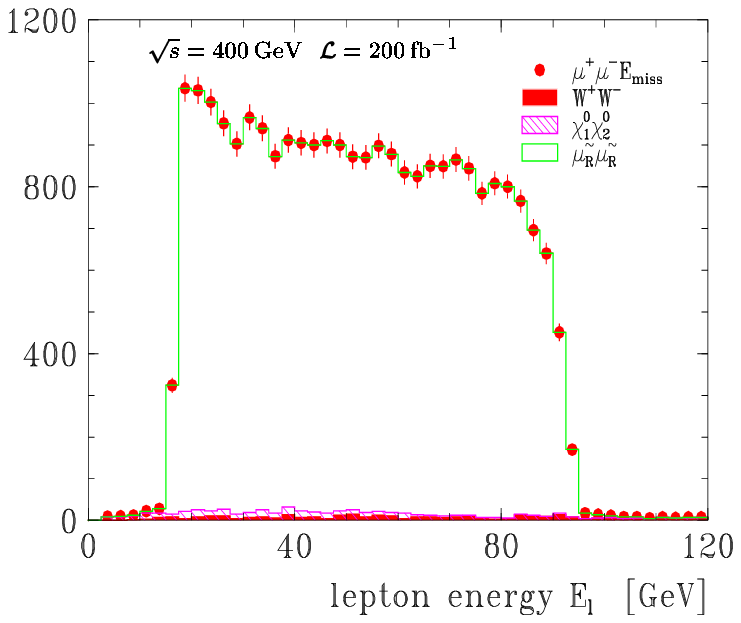}
\caption{Left: Mass measurement in chargino $\tilde{\chi}^+_1\tilde{\chi}^-_1$
         pair production; Right: Smuon and neutralino edges in two-body
         smuon decays,
         $\tilde{\mu}^\pm_R\to\mu^\pm\tilde{\chi}^0_1$;
         Ref\,\cite{R23}.}
\label{fig7}       
\end{figure}
\begin{table}
\caption{Accuracies for representative mass measurements of SUSY particles in
         individual LHC, ILC and coherent LHC/ILC analyses for the point
         SPS1a$'$ [mass units in GeV]; Ref.\,\cite{R10}.}
\label{tab:1}   
\vskip 0.3cm
\begin{center}
\begin{tabular}{|c|c||c|c||c|}
\hline
Particles & Mass &  ``LHC"  & ``ILC"  & ``LHC+ILC" \\
\hline \hline
$h^0$                 & 116.0 & 0.25 & 0.05 & 0.05 \\
$H^0$                 & 425.0 &      & 1.5  & 1.5  \\
    \hline
$\tilde{\chi}^0_1$    &  97.7 & 4.8  & 0.05 & 0.05 \\
$\tilde{\chi}^0_2$    & 183.9 & 4.7  & 1.2  & 0.08 \\
$\tilde{\chi}^0_4$    & 413.9 & 5.1  & 3-5  & 2.5  \\
$\tilde{\chi}^\pm_1$  & 183.7 &      & 0.55 & 0.55 \\ \hline
$\tilde{e}_R$         & 125.3 & 4.8  & 0.05 & 0.05 \\
$\tilde{e}_L$         & 189.9 & 5.0  & 0.18 & 0.18 \\
$\tilde{\tau}_1$      & 107.9 & 5-8  & 0.24 & 0.24 \\ \hline
$\tilde{q}_R$         & 547.2 & 7-12 & -    & 5-11 \\
$\tilde{q}_L$         & 564.7 & 8.7  & -    & 4.9  \\
$\tilde{t}_1$         & 366.5 &      & 1.9  & 1.9  \\
$\tilde{b}_1$         & 506.3 & 7.5  & -    & 5.7  \\ \hline
$\tilde{g}$           & 607.1 & 8.0  & -    & 6.5  \\ \hline
\end{tabular}
\end{center}
\end{table}

At ILC very precise mass values can be extracted from threshold scans and decay
spectra \cite{R23}. The excitation curves for chargino $\tilde{\chi}^\pm_{1,2}$
production in S-waves rise steeply with the velocity of the particles near
threshold and they are thus very sensitive to the mass values, the left panel
of Fig.\,\ref{fig7}. The same holds true for mixed chiral selectron pairs in
$e^+e^-\to\tilde{e}^+_R\tilde{e}^-_L$ and for diagonal pairs in
$e^-e^-\to \tilde{e}^-_R\tilde{e}^-_R, \tilde{e}^-_L\tilde{e}^-_L$ \cite{R24a,R24b}.
Other scalar fermions as well as neutralinos are produced in P-waves
with a less steep threshold behavior proportional to the third
power of the velocity. An important information on the mass of the LSP
such as the lightest neutralino $\tilde{\chi}^0_1$ can be obtained
from the sharp edges of two-body decay spectra such as $\tilde{\ell}_R
\to \ell \tilde{\chi}^0_1$, the right panel of Fig.\,\ref{fig7} \cite{R23}.
%
%
The accuracy in the measurement of the LSP mass can be improved at ILC by
two orders of magnitude compared with LHC; Tab.\,\ref{tab:1}.


The values of typical mass parameters and their related measurement errors are
presented in Tab.\,\ref{tab:1}: ``LHC" from LHC analyses and ``ILC" from
ILC analyses. The fourth column ``LHC+ILC" represents the corresponding
errors if the experimental analyses are performed coherently \cite{R10}.

\begin{figure}[h]
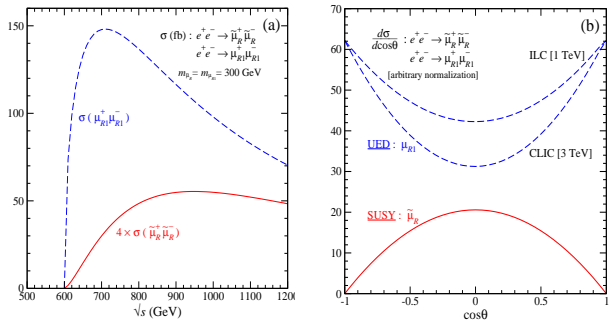

\vskip 0.0cm
\begin{center}
\includegraphics[width=0.23\textwidth,height=0.246\textwidth,angle=0]{smuon_threshold.eps}
\hskip 0.3cm
\includegraphics[width=0.22\textwidth,height=0.249\textwidth,angle=0]{smuon_angle.eps}
\end{center}
\caption{The threshold excitation (a) and the angular distribution (b) in the
         case of smuons in the MSSM and the first Kaluza-Klein muons in an adopted
         model of universal extra dimensions in pair production at ILC; For details,
         see Ref.\,\cite{R26b}.}
\label{fig8}       
\end{figure}

Determining the spin of new particles is an important method to clarify the
nature of the particles and the underlying theory. This determination is
crucial to distinguish the supersymmetric interpretation of new particles
from other models. The measurement of the spins in particle cascades at LHC
is quite involved \cite{R25a,R25b,R25c}. In contrast spin measurement at ILC is
straightforward \cite{R26a,R26b}. The $\sin^2\theta$ law for the angular
distribution in the production of sleptons (for selectrons close to threshold)
is a unique signature of the fundamental spin-zero character; the P-wave onset
of the excitation curve is a necessary but not sufficient condition;
Fig.\,\ref{fig8}. In contrast, neither the onset of the excitation curves near
threshold nor the angular distribution in the production processes provide unique
signals of the spin of charginos and neutralinos. However, decay angular
distributions of polarized charginos/neutralinos  could provide an
unambiguous determination of the spin-1/2 character of the particles
albeit at the expense of more involved experimental analyses \cite{R26b}.
[Quantum interference among helicity amplitudes, reflected on azimuthal angle
distributions, may also be used to determine spin in a model-independent
way \cite{azimuthal}.]

\begin{figure}[h]
\vskip 0.0cm
\begin{center}
\includegraphics[width=0.35\textwidth,height=0.3\textwidth,angle=0]{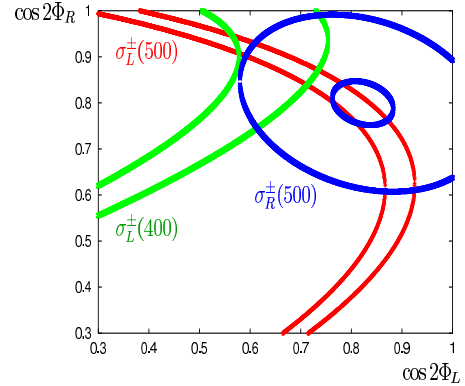}
\end{center}
\caption{Contours for the $e^+e^-\to\tilde{\chi}^+_1\tilde{\chi}^-_1$ production
         cross section for polarized $e^\pm$ beams at $\sqrt{s}=400$ and $500$
         GeV; Ref.\,\cite{R27c}.}
\label{fig9}       
\end{figure}

Mixing parameters must be extracted from measurements of cross sections and
polarization asymmetries. In the production of charginos and neutralinos,
both diagonal and mixed pairs can be exploited: $e^+e^-$ $\to$ $\tilde{\chi}^+_i
\tilde{\chi}^-_j$ [$i,j=1,2$] \cite{R27a,R27b,R27c} and
$\tilde{\chi}^0_i\tilde{\chi}^0_j$ [$i,j=1,..,4$] \cite{R28}. The production
cross sections for charginos are binomials in $\cos 2\phi_{L,R}$ where
$\phi_{L,R}$ are the mixing angles rotating current to mass eigenstates. Using
polarized electron and positron beams, the mixings can be determined in a
model-independent way, Fig.\,\ref{fig9}. The same procedures can be applied to
determine the mixings in the sfermion sector \cite{R29a,R29b,R29c}.
The production cross sections for stop particle pairs,
$e^+e^-\to\tilde{t}_i \tilde{t}^*_j$ [$i,j=1,2$], depend on the
stop mixing angle $\theta_{\tilde{t}}$ which can be determined
with high accuracy by use of polarized electron beams \cite{R29c}.

SUSY predicts the identity of Yukawa and gauge couplings among supersymmetric
partners for gauge bosons and gauginos, and for fermions and their scalar partners.
On one hand the fundamental SU(3)$_C$ relation can be studied experimentally
at LHC through pair production of squarks partly mediated by gluino $t$-channel
exchanges \cite{R30a,R30b}. On the other hand the SU(2)$_L$ and U(1)$_Y$ relations can be
confirmed experimentally at ILC through pair production of charginos and neutralinos
which is partly mediated by the exchange of sneutrinos and selectrons in the
$t$-channel \cite{R28}, as well as selectron and sneutrino production which is
partly mediated by neutralino and chargino exchanges \cite{R24b}.
The separation of the electroweak SU(2) and U(1) couplings is also possible
if polarized electron beams are available. Of course the analysis for confirming
the identity of Yukawa and gauge couplings should be
performed by taking into account the prior measurements of the masses and/or mixing
parameters of the particles exchanged in the $t$-channel.

\subsection{Fundamental theory}
\label{sec:4-2}

Combining the information from LHC on the generally heavy colored particles
with the information from ILC on the generally lighter non-colored particle
sector (and later from CLIC on heavier states) will generate a model-independent
and high-precision picture of SUSY at the Terascale. The picture may subsequently
serve as a solid platform for the reconstruction of the fundamental SUSY theory
at a high scale, potentially close to the Planck scale, and for the analysis
of the microscopic mechanism of SUSY breaking \cite{R31,Raby}. The experimental
accuracies expected at the percent down to the per-mil level must be matched on
the theoretical side. This demands a well-defined framework for the calculational
schemes in perturbation theory as well as for the input parameters like
a recently proposed scheme called Supersymmetry Parameter Analysis (SPA) \cite{R10}.

If SPS1a$^\prime$ or a similar SUSY parameter set is realized in nature, various
channels can be exploited to extract the basic Terascale SUSY parameters at LHC
and ILC. The data analysis performed coherently for LHC and ILC is shown to
give rise to a very precise picture of the supersymmetric particle spectrum.
Running global analysis programs with the whole set of data enables us to
coherently extract the Lagrangian parameters in the optimal way after including
the available radiative corrections \cite{R32a,R32b,R32c,R32d,R32e}.
The present quality of such an analysis can be judged from the results shown
in Tab.\,\ref{tab:2}.

%
\begin{table}
\caption{Excerpt of extracted SUSY Lagrangian mass and Higgs parameters at the
         Terascale in the reference point SPS1a$'$ [mass units in
         GeV]; Ref.\,\cite{R10}.}
\label{tab:2}   
\vskip 0.3cm
\begin{center}
\begin{tabular}{|c|c|c|}
\hline
Parameter & SPS1a$'$ value &  Fit error [exp] \\
\hline \hline
$M_1$             & 103.3 & 0.1  \\
$M_2$             & 193.2 & 0.1  \\
$M_3$             & 571.7 & 7.8  \\
$\mu$             & 396.0 & 1.1  \\
    \hline
$M_{L_1}$    & 181.0 &  0.2 \\
$M_{E_1}$    & 115.7 &  0.4 \\
$M_{L_3}$    & 179.3 &  1.2 \\
   \hline
$M_{Q_1}$    & 525.8 &  5.2 \\
$M_{D_1}$    & 505.0 & 17.3 \\
$M_{Q_3}$    & 471.4 &  4.9 \\
   \hline
$m_A$          & 372.0  & 0.8 \\
$A_t$          & -565.1 & 24.6 \\
$\tan\beta$    & 10.0 & 0.2\\ \hline
\end{tabular}
\end{center}
\end{table}

Based on the parameters extracted at the Terascale we can reconstruct the
fundamental SUSY theory and the related microscopic picture of the SUSY
breaking mechanism \cite{R31}. The experimental information is exploited
to the maximum extent possible in the bottom-up approach in which the
extrapolation from the Terascale to the GUT/Planck scale is performed
by the renormalization group (RGE) evolution of all parameters, with the
GUT scale defined by the unification point of the two electroweak couplings.

\begin{figure}[h]
\begin{center}
\includegraphics[width=0.47\textwidth,height=0.25\textwidth,angle=0]{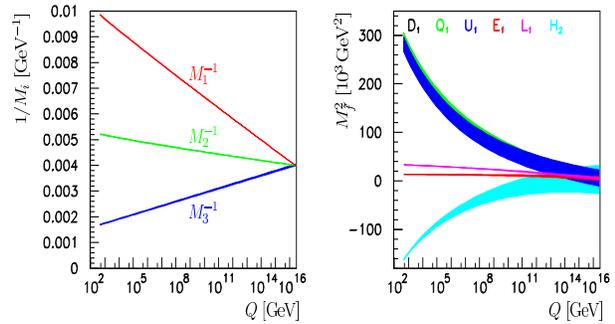}
\end{center}
\caption{Running of the gaugino and scalar mass parameters as a function of the
         scale $Q$ in SPS1a$^\prime$. Only experimental errors are taken into
         account; theoretical errors are assumed to be reduced to the same size
         in the future; Ref.\,\cite{R31}.}
\label{fig10}       
\end{figure}

Typical examples for the evolution of the gaugino and scalar mass
parameters are presented in Fig.\,\ref{fig10}. While the determination
of the high-scale parameters in the gaugino and higgsino sector, as well
as in the non-colored slepton sector, is very precise, the picture of
the colored scalar and Higgs sectors is still coarse so that considerable
efforts should be made to refine it considerably.
If the structure of the theory at the GUT scale was known a priori and merely
the experimental determination of the high scale parameters were lacking, then
the top down-approach would lead to a very precise parametric picture at the
Terascale.

So far, we have only considered the MSSM, in particular the parameter set
SPS1a$'$, as a benchmark scenario for judging the coherent capabilities of LHC
and ILC experiments for a successful analysis of future SUSY data. However,
neither this specific point nor the MSSM itself may be the correct model for
low-scale SUSY. Various extended models have therefore to be investigated.
In particular, models which incorporate the right-handed neutrino sector to
accommodate the complex structure observed in the neutrino sector must be
scrutinized in a systematic way \cite{R33}. Furthermore, CP violation,
R-party violation \cite{R34}, flavor violation \cite{R35}, NMSSM \cite{R36a,R36b}
and/or extended gauge groups \cite{R20,R37} are among the paths that nature may
have taken. It is, therefore, strongly recommended that the analysis conventions
and methods be so general that they can be applied to all those BSM
scenarios as well.

\subsection{SUSY dark matter}
\label{sec:4-3}

Since there is no proper CDM particle candidate in the SM,
the presence of the CDM is a clear evidence for physics beyond the SM.
In SUSY theories with R-parity the LSP is absolutely stable and represents a
good CDM candidate \cite{R8}. In particular, the lightest neutralino is
considered to be the prime candidate, but other interesting possibilities are the
axino and the gravitino.

\begin{figure}[h]
\begin{center}
\includegraphics[width=0.45\textwidth,height=0.25\textwidth,angle=0]{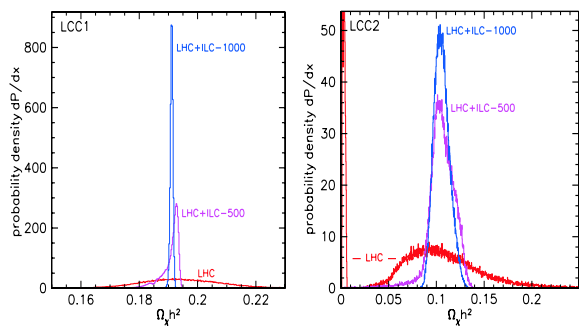}
\end{center}
\caption{Probability distribution of predictions of $\Omega_\chi h^2$ for
         the LCC1 ``bulk" point and the LCC2 ``focus-point" point from
         measurements at ILC with $\sqrt{s}=0.5$ and $1$ TeV, and LHC (after
         qualitative identification of the model); Ref.\,\cite{R38}}
\label{fig11}       
\end{figure}

In certain areas of the SUSY parameter space with the $\tilde{\chi}^0_1$ relic
density in the range required by WMAP, SUSY particles can be produced
abundantly at LHC and ILC. However, to determine the predicted WMAP
relic density, we must have detailed knowledge not only of the LSP
properties but also of all other particles contributing to the
LSP pair annihilation cross section. To quantify the prospects for
determining the neutralino CDM relic density at ILC as well as LHC,  and
the connection of ILC with cosmology, four benchmark mSUGRA scenarios called LCC
points and compatible with WMAP data have been proposed \cite{R38}.
The ILC measurements at $\sqrt{s}=0.5$ TeV and $1$ TeV for various
sparticle masses and mixings in the scenarios, taking into account
LHC data, are compared to those which can be obtained using LHC data
(after a qualitative identification of the model). As can be seen in
Fig.\,\ref{fig11} for two LCC points, the LCC1 ``bulk" point and the LCC2
``focus-point" point, the gain in sensitivity by combining LHC and ILC
is spectacular.

In supergravity models the gravitino $\tilde{G}$ itself may be the LSP,
building up the dominant CDM component \cite{R39a,R39b,R39c,R39d}. In such a scenario, with a
gravitino mass in the range of 100 GeV, the lifetime of the next-to-LSP (NLSP) can become
macroscopic as the gravitino coupling is only of gravitational
strength. The lifetime of the NLSP $\tilde{\tau}$ can extend up to several
months, suggesting special experimental efforts to catch the long-lived
$\tilde{\tau}$'s and to measure their lifetime \cite{R40a,R40b,R40c,R40d}. Tau slepton pair
production at ILC determines the $\tilde{\tau}$ mass and the observation of the
$\tau$ energy in the $\tilde{\tau}$ decay determines the gravitino mass.
The measurement of the lifetime can subsequently be exploited to
determine the Planck scale, a unique opportunity in a laboratory
experiment.

\section{Conclusions}
\label{sec:4}

The next generation of high energy experiments, LHC and ILC (and also CLIC later),
will usher us into the Terascale, opening a new territory which is highly expected
to be full of ground-breaking discoveries. The physics programme of both LHC and
ILC in exploring this microscopic world will be very rich with their unique
characteristics depending on the BSM physics scenario realized in nature.
Furthermore, as demonstrated by dedicated studies using the SUSY models, the
physics potential of LHC and ILC can significantly be extended by coherent or/and
``concurrent" running of both machines.

In summary, the LHC and ILC experiments with different advantages and
capabilities can contribute coherently and complementarily to solutions of
key questions in particle physics and cosmology. Eventually both experiments
can provide us with a comprehensive and high-resolution picture not only of
SUSY but also of any alternative scenario, serving as a telescope to
unification of matter and interactions, and connection of particle physics
and cosmology.

\subsection*{Acknowledgements}
\label{sec:ack}

This work was supported in part by the Korea Research Foundation Grant funded by
the Korean Government (MOEHRD, Basic Research Promotion Fund)
(KRF-2007-521-C00065) and in part by KOSEF through CHEP at Kyungpook National
University.

%
%

\end{document}